\documentclass[12pt]{article}
\usepackage{geometry}                      
\usepackage{graphicx}
\usepackage{amssymb}
\usepackage{natbib}
\setcitestyle{aysep={,}}
\usepackage{amsmath}   
\usepackage{bm}
\usepackage{setspace}
\usepackage{url}

\begin{document}

\title{Teaching Bayes' Rule using Mosaic Plots}

\author{
Edward D. White\footnote{Department of Mathematics and Statistics, Air Force Institute of Technology, Wright-Patterson Air Force Base, Ohio, USA.  The views expressed in this article are those of the authors and do not reflect the official policy or position of the United States Air Force, Department of Defense, or the U.S. Government.}  \, and 
Richard L. Warr\footnote{Department of Statistics, Brigham Young University, Provo, Utah, USA} }
\maketitle

Students taking statistical courses orientated for business or economics
often find the standard presentation of Bayes' Rule challenging. This key
concept involves understanding multiple conditional probabilities and how
they constitute an unconditional sample
space. Many textbooks try to aid the comprehension of Bayes' Rule by
illustrating these probabilities with tree diagrams. In our opinion, these
diagrams fall short in fully assisting the students to \textit{visualize}
Bayes' Rule. In this article, we demonstrate a graphical
approach that we have successfully used in the classroom, but is neglected in introductory texts. This approach uses mosaic plots to show the weighting of the conditional probabilities and greatly
aids the student in understanding the sample space and its associated probabilities.


\newpage

\section{Introduction}

Often statisticians hear from acquaintances that statistics was their worst subject in college. Collectively the statistics community is trying to correct this problem. Part of the solution is to improve teaching methods.

Most people are visual learners \citep{hattie2012visible,riding2013cognitive}. 
Some reports have cited that more than 80\% of human learning occurs visually. 
Additionally, \cite{newcombe2010picture}  finds that spatial thinking is vital to understanding math and science. Historically, most college teaching has been communicated verbally: ``the information presented is predominantly auditory (lecturing) or a visual representation of auditory information (words and mathematical symbols written in texts and handouts, on transparencies, or on a chalkboard)'', \cite{felder1988learning}.  Although classroom presentation style has been gradually shifting over the years, information is still predominately presented in the same way.  Additionally, \cite{simmons2014effective} and \cite{moreno2011teaching} suggest linking visual and numerical conceptualizations to improve student problem-solving skills.

Having taught business statistics for several years, we have realized Bayes' Rule initially intimidates most students. This is unfortunate given the importance of Bayes' Rule. The power of Bayes' Rule comes from reversing the conditional variables. It is our opinion that students struggle with the typical formulation of Bayes' Rule, especially computing the denominator using the law of total probability. From our experience we suggest a mosaic plot approach to visually represent the typical mathematical presentation of Bayes' Rule.

\section{Mosaic Plot Approach}

Equation \ref{eq1} is a typical mathematical representation of Bayes' Rule, as found in \cite{mcclave2011statistics} and \cite{newbold2010statistics} (and many other texts). Conditional probabilities themselves can easily trip up introductory students and Bayes' Rule adds to that complexity.

\begin{equation}
P\left(B_i|A\right) = \frac{P\left(A|B_i\right)P\left(B_i\right)}{
\sum_{j=1}^{k}P\left(A|B_j\right)P\left(B_j\right)}.  \label{eq1}
\end{equation}

Based on our experience, strictly formulating Bayes' Rule mathematically seems to hinder many students from comprehending this fundamental concept. This is unacceptable, given the real-world applicability of Bayes' Rule.  Examples of how Bayes' Rule is used on a daily basis are numerous, for instance, calculating the chance one actually has cancer, given a positive indication from a cancer screening test. Or, finding the probability an uninsured motorist has struck your car, given you were involved in a car accident. 

\begin{figure}[ht]
\begin{center}
\includegraphics[width=2.5in]{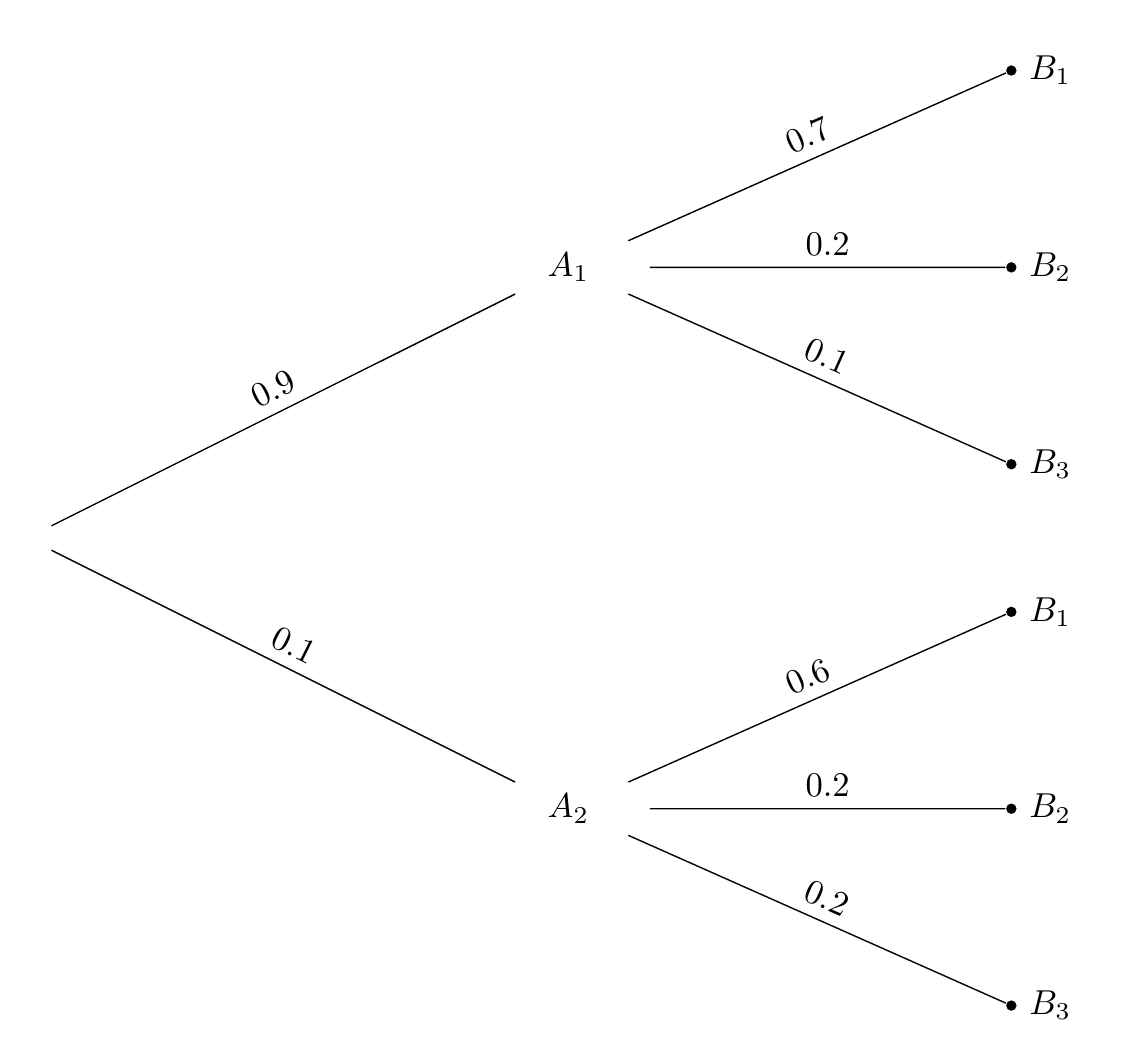}
\end{center}
\caption{Typical representation of a probability tree whereby the
probability of one event, $B$, changes depending on whether another event, $A
$, occurred.}
\label{fig1}
\end{figure}

To assist those students who prefer a visual presentation of material, some textbooks accompany and augment Bayes' Rule by tree diagrams. Figure \ref{fig1} represents a typical example of a tree diagram. These tree diagrams illustrate the unconditional probabilities as main branches stemming from the tree and then conditional probabilities representing branches from these main trunk stems.

Although tree diagrams help some students comprehend Bayes' Rule it does not visually depict the weighting of the probabilities on each branch.  In Figure  \ref{fig1}, we see two main branches representing event $A_1$ or $A_2$. From these branches, the three sub-branches represent the events $B_1$, $B_2$, or $B_3$. Pictorially these three sub-branches appear symmetric, when in fact their associated probabilities are considerably different. It is not until students more deeply grasp the concept that they understand the relative weighting of these conditional probabilities.

We advocate a visual method to introduce Bayes' Rule.  While ample research over the past few decades has examined the errors people make in performing statistical reasoning, a lesser but growing body of research has addressed how to teach people to solve statistical problems \citep{gigerenzer1999overcoming,kurzenhauser2002teaching,sedlmeier2000improve}. For conditional probability problems, \cite{kellen2013improving} suggests keeping the user's focus on a limited number of elements. Additionally, they use visual elements in lieu of text wherever possible, thus reducing the extraneous cognitive load. This can be done by giving the user perceptual cues such as shading, color or sequencing to focus their attention on a small number of elements at a time. This approach helps people build a partial mental model and integrate the partial model into a full model where the correct inference can then be made \citep{pollock2002assimilating}.

With this in mind, we adopt a slight modification of the mosaic plots usually attributed to \cite{hartigan1981mosaics, hartigan1984mosaic}, although \cite{friendly2002brief} traced these plots back to even earlier work. As defined by \cite{friendly2000visualizing},``the mosaic display is a graphical method for visualizing an n-way contingency table and for building models to account for the association among its variables. The frequencies in a contingency table are portrayed as a collection of rectangular `tiles' whose areas are proportional to the cell frequencies.''

In our use of mosaic displays, we use the standard rectangular approach that reflects the starting sample space, and partitions this space into the respective unconditional and conditional probabilities. The unconditional parts will be described as the \textit{smooth} rectangles, while the conditional parts will be described as the \textit{jagged} rectangles (this is visualized shortly).  With this method, students see the weighting of the probabilities, and also how each piece is related to the others. Some concepts for this approach are found in \citep[pg. 454]{wilkinson2006grammar}, but have not been integrated into mainstream introductory statistical teaching literature.

This graphical approach builds on the fact that Bayes' Rule is based on events that are mutually exclusive and exhaustive. In other words, the sample space, $S$, consists of a collection of smaller rectangles that do not overlap but in totality fill $S$. We illustrate this method of visually presenting Equation \ref{eq1} through two examples. The first example serves as a typical presentation of Bayes' Rule. Our second example is more complicated and presents a $4 \times 4$ problem, which is challenging to present visually by a tree diagram approach.

\subsection{Example 1}

For our first example, we essentially duplicate one given by \cite[pg. 159]{mcclave2011statistics}. An unmanned surveillance system is designed to successfully detect an intruder with probability $P(A_1)$ and a failure probability of $P(A_2)$, with $A_1$ and $A_2$ being mutually exclusive and exhaustive events. Based upon whether an intruder is successfully detected or not, we also record the weather during this occurrence. Call these weather events $B_1$, $B_2$, or $B_3$ for clear, cloudy, or rainy respectively (also mutually exclusive and exhaustive). The decision tree for this example has two main branches, $A_1$ or $A_2$, and then three sub-branches for $B_1$, $B_2$, and $B_3$ on each main branch. Figure \ref{fig1}, as illustrated earlier, represents the tree diagram for this example, while Table \ref{tab1} presents the unconditional and conditional probabilities.

\begin{table}[ht]
\caption{The probabilities found in Figure \protect\ref{fig1}.}
\label{tab1}
\begin{center}
\begin{tabular}{|c|c|c|c|}
\hline
Unconditional Probabilities & \multicolumn{3}{|c|}{Conditional Probabilities}
\\ \hline
$P(A_1)=0.90$ & $P(B_1|A_1)=0.70$ & $P(B_2|A_1)=0.20$ & $P(B_3|A_1)=0.10$ \\ 
$P(A_2)=0.10$ & $P(B_1|A_2)=0.60$ & $P(B_2|A_2)=0.20$ & $P(B_3|A_2)=0.20$ \\ 
\hline
\end{tabular}%
\end{center}
\end{table}

\begin{figure}[ht]
\begin{center}
\includegraphics[width=2.5in]{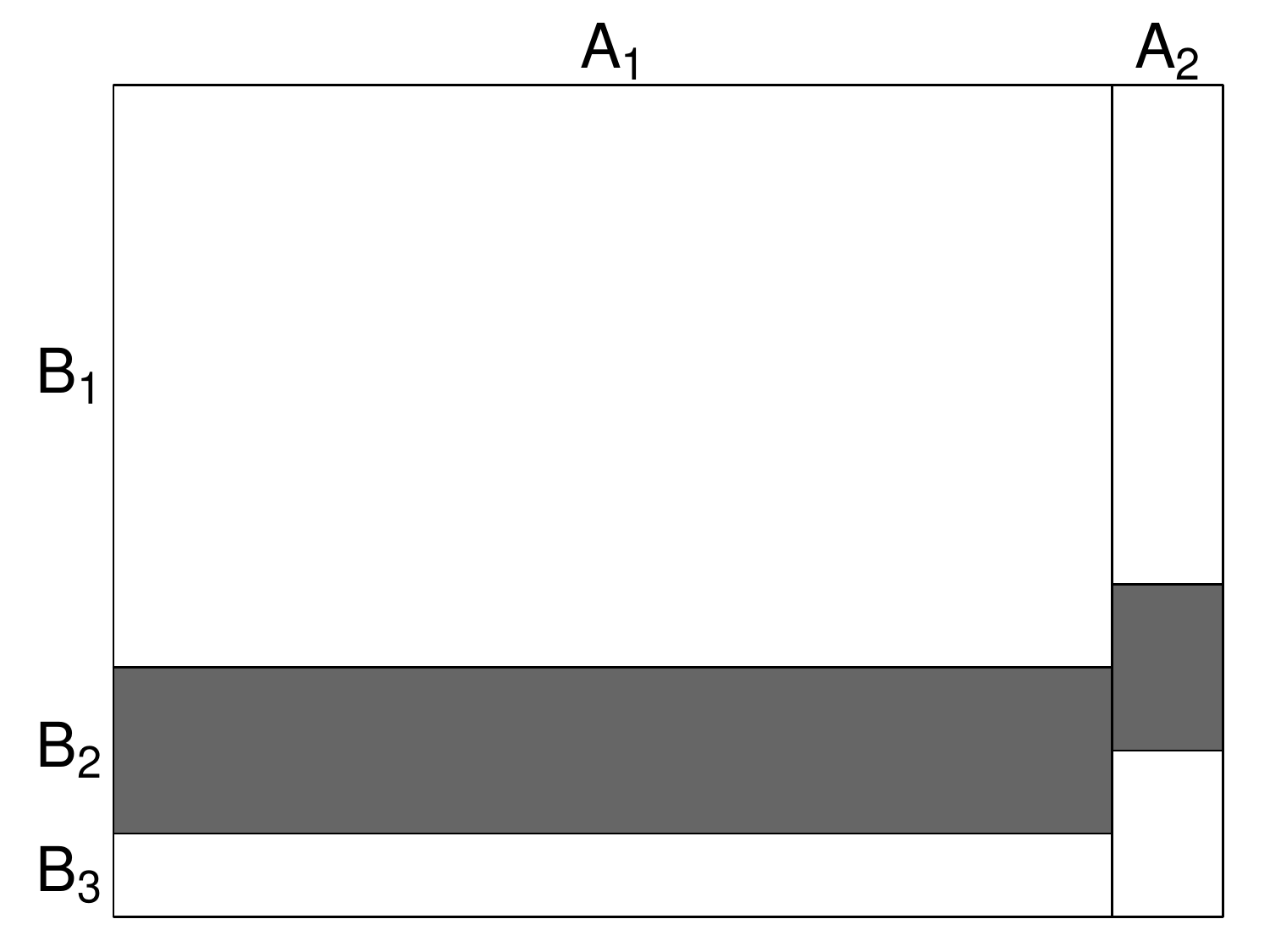}
\end{center}
\caption{Probability mosaic plot representation of the probability tree
presented in Figure \protect\ref{fig1}. The shaded area represents the
probability of $B_2$ occurring regardless of event $A$.}
\label{fig3}
\end{figure}

Given these probabilities, suppose we wish to answer the question, ``What is the probability that this unmanned surveillance system will detect an intruder given it's cloudy?'' Figure \ref{fig3} represents the use of a mosaic plot to not only visually augment the mathematical formulation of Bayes' Rule as shown in Equation \ref{eq1}, but to also help display the weighting of the probabilities as given by the problem. The \textit{smooth} rectangles of $A_1$ and $A_2$ represent the unconditional elements of our example, while the \textit{jagged} rectangles of $B_1$, $B_2$, and $B_3$ represent the conditional elements of our example.

The entire rectangle in Figure \ref{fig3} represents the complete sample space for our example. All of the boxes within this sample space represent all the possibilities. Consequently the total area of all those little rectangles must sum to one in accordance with the axioms of probability. The shaded area in Figure \ref{fig3}, however, represents just the area or probability of $B_2$, which for this problem represents the event that it is cloudy. There are two rectangular pieces to this event: the piece where it is cloudy and the system detected an intruder and the other part when it is cloudy and the system did not detect the intruder. If we took these two shaded boxes, we have the total piece that represents the area or probability that it is cloudy. This is the bottom part of Equation \ref{eq1}. For the top part of Equation \ref{eq1}, we recognize that this numerator is just the piece of the total \textit{cloudy} rectangle that represents just the part where it is cloudy and the system detected the intruder.

\begin{figure}[ht]
\begin{center}
\includegraphics[width=2.5in]{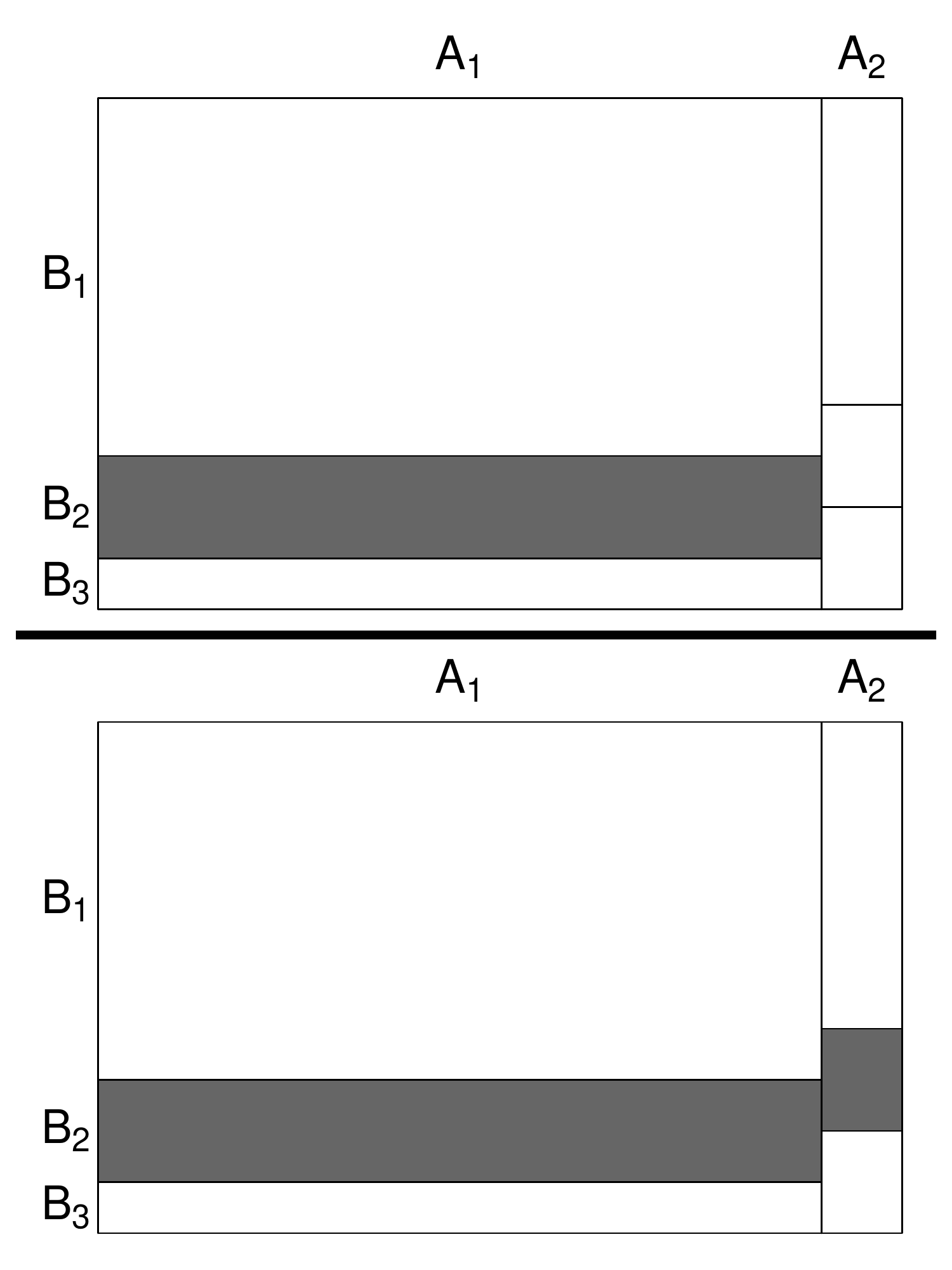}
\end{center}
\caption{Ratio of two probability mosaic plots. The plot in the numerator
highlights the event $A_1 \cap B_2$. The plot in the denominator shades the
event $B_2$ (unconditionally) and is graphed separately in Figure \protect
\ref{fig3}.}
\label{fig4}
\end{figure}

So to answer the original question (now framed in a mathematical formulation), $P(A_1|B_2)$, we can visualize Equation \ref{eq1} as being a ratio of probability rectangles, whereby the top rectangle is just one element of shaded portion in Figure \ref{fig3}. Figure \ref{fig4} presents the mosaic plot of $P(A_1|B_2)$.

In our classes, we have found this visualizing of Equation \ref{eq1} to be quite useful. In this regard, it is easy to convey that our entire sample space represents a box of puzzle pieces that when dumped out on a table can be put together to form one large rectangle. Once we state the conditioning event, the student can then take all those pieces from the sample space that contain that conditioning event and those pieces now form their own puzzle. The numerator of Equation \ref{eq1} is now simply one of those pieces from the new puzzle the student has put together. Adding color to these puzzle pieces corresponding to the conditioning events almost always get the students to immediately see the various conditioning events and, in our experiences, quickly understand and use Equation \ref{eq1}. 

\subsection{Example 2}

\begin{figure}[ht]
\begin{center}
\includegraphics[width=2.5in]{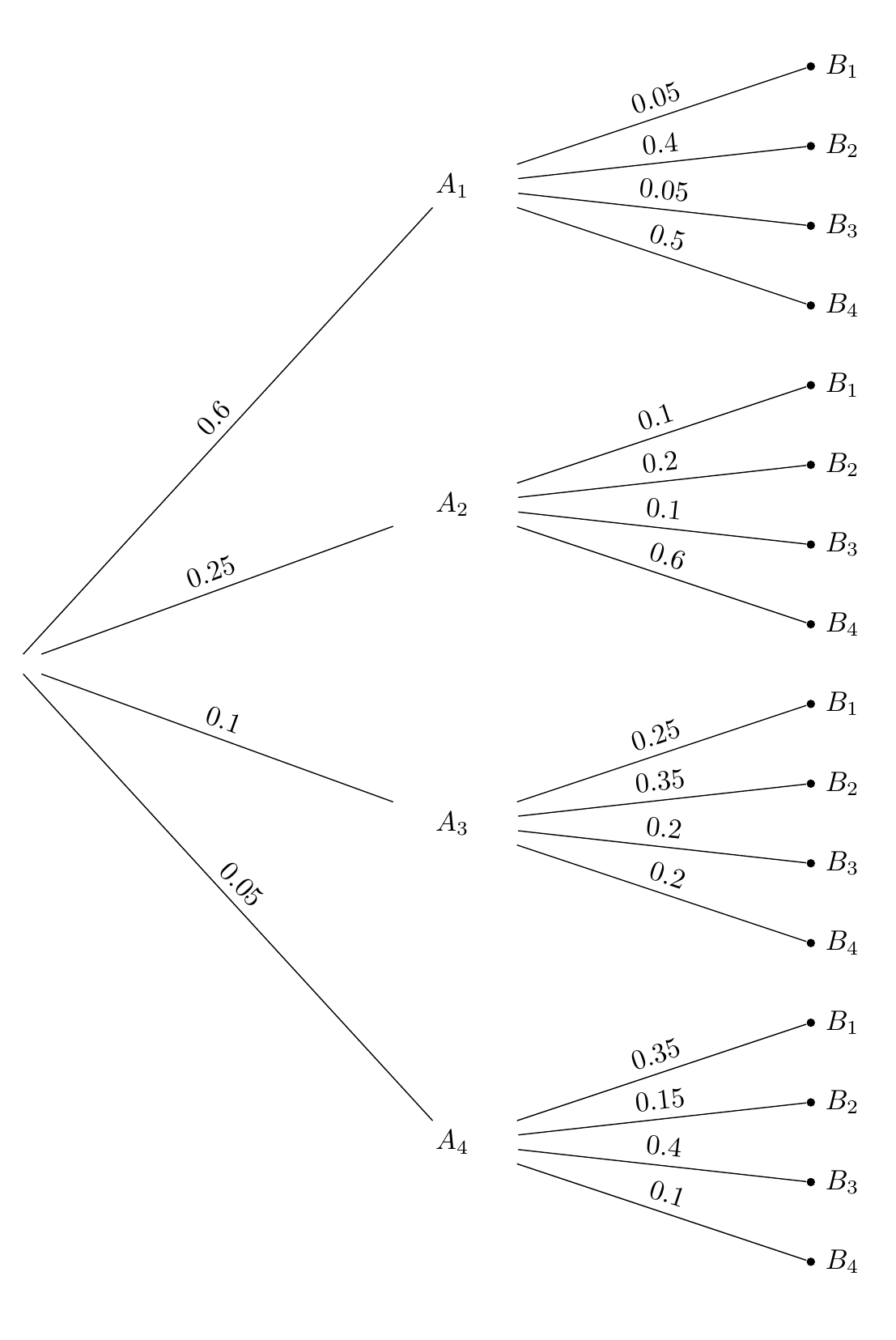}
\end{center}
\caption{Probability tree displaying the unconditional probabilities for
event $A$ and the conditional probabilities for event $B$ given $A$ in the $
4 \times 4$ scenario.}
\label{figTree4x4}
\end{figure}

The other example represents a $4 \times 4$ scenario and demonstrates how the typical probability tree diagram approach is less effective. In this example, event $A$ has four possibilities and another four with event $B$ conditioned on $A$. For simplicity, we make the assumption that all four levels of event $B$ are possible in each partition of $A$. Table \ref{tab3} relays an example of the unconditional and conditional probabilities of events $A$ and $B$, while Figure \ref{figTree4x4} contains the tree diagram.

\begin{table}[ht]
\caption{Unconditional and conditional probabilities for the $4 \times 4$
scenario. There are 4 possibilities for event $A$, and for each possibility
of $A$ there are four more possibilities for $B|A$.}
\label{tab3}
\begin{center}
\begin{tabular}{|c|c|c|c|c|}
\hline
Uncond Probs & \multicolumn{4}{|c|}{Conditional Probabilities} \\ \hline
$P(A_1)=0.60$ & $P(B_1|A_1)=0.05$ & $P(B_2|A_1)=0.40$ & $P(B_3|A_1)=0.05$ & $
P(B_4|A_1)=0.50$ \\ 
$P(A_2)=0.25$ & $P(B_1|A_2)=0.10$ & $P(B_2|A_2)=0.20$ & $P(B_3|A_2)=0.10$ & $
P(B_4|A_2)=0.60$ \\ 
$P(A_3)=0.10$ & $P(B_1|A_3)=0.25$ & $P(B_2|A_3)=0.35$ & $P(B_3|A_3)=0.20$ & $
P(B_4|A_3)=0.20$ \\ 
$P(A_4)=0.05$ & $P(B_1|A_4)=0.35$ & $P(B_2|A_4)=0.15$ & $P(B_3|A_4)=0.40$ & $
P(B_4|A_4)=0.10$ \\ \hline
\end{tabular}
\end{center}
\end{table}

\begin{figure}[ht]
\begin{center}
\includegraphics[width=2.5in]{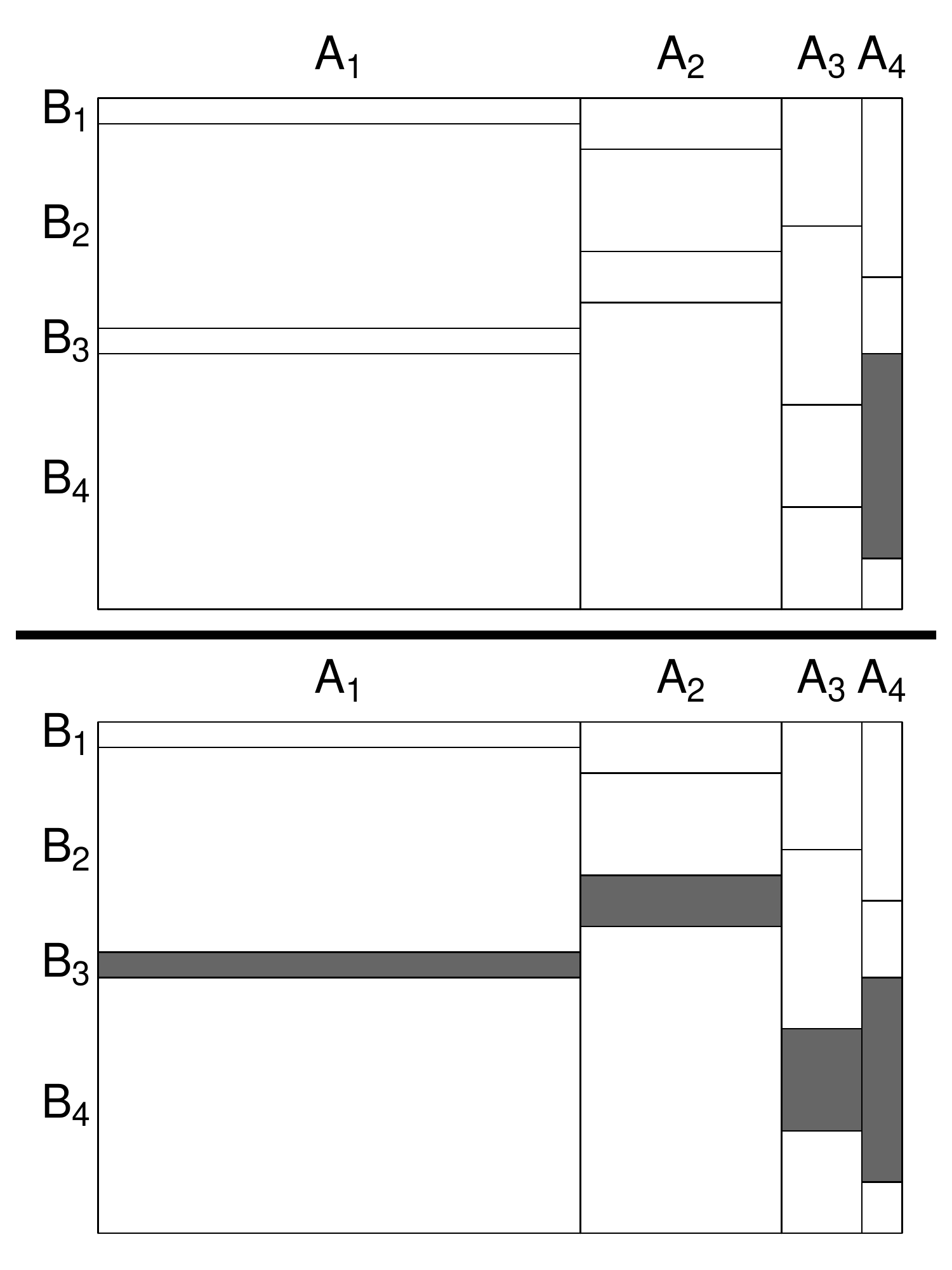}
\end{center}
\caption{Visual representation of the $4 \times 4$ example using a ratio of
probability mosaic plots. Shaded regions in the denominator represent the
event $B_3$ and the shaded region in the numerator represents the event $A_4
\cap B_3$.}
\label{fig4x4}
\end{figure}

Using the mosaic plot approach, assume that event $B_3$ has occurred. This event's probability is visualized by the collection of shaded rectangles in the denominator of the mosaic probability plot in Figure \ref{fig4x4}. Given $B_3$ occurred, what is the probability event $A_4$ occurred? This probability is represented by the area of the single shaded rectangle in the numerator of the mosaic plot in Figure \ref{fig4x4}. Clearly the shaded rectangle in the numerator is just one of the four shaded rectangles in the denominator. Before calculating the answer via Equation \ref{eq1}, a student can visualize the solution to the question as the ratio of the shaded areas. This is not a trivial problem since we are dealing with a $4 \times 4$ scenario here.

The mosaic plots in Figure \ref{fig4x4} could easily be modified for a $5 \times 5$ or an $n \times m$ scenario. Our students readily followed the puzzle pattern when the mathematical formulation of Bayes' Rule in Equation \ref{eq1} was visually augmented in this fashion. Students visualized the answer before tackling the math of Equation \ref{eq1}. This graphical approach, in our experience, facilitated learning where the mathematics became a building block and not a stumbling block.

\section{Conclusion}

In introductory statistical courses, most students find the subject of probability, and in particular the learning of Bayes' Rule, challenging especially when presented mathematically as in Equation \ref{eq1}. What generally needs to be added to accommodate these students is visual material—pictures, diagrams, or sketches. We concur that ``mathematical functions should be illustrated by graphs'' \cite{felder1988learning}. 

Tree diagrams are usually the norm for augmenting the instruction of Bayes' Rule for they can easily show the ordering of events. But like most tree diagrams, they do not show the weighting of the conditional and unconditional probabilities. A teacher can also use or construct a table of frequencies. This is in keeping with the Gigerenzer school of thought that people are much better with frequencies than with probabilities. But, in our experience, this limits the visualization of Bayes' Rule for students do not \textit{see} the probabilities and may have a difficult time understanding where to put the numbers like in Equation \ref{eq1}.

Using the mosaic plot approach to present Bayes' Rule allows students to see the proper weighting of the conditional and unconditional probabilities, but also allows students to visualize precisely the question that is being asked, and the construction of the answer. This mosaic probability plot can easily be extended to a multitude of events and surpasses the visual clarity of a tree diagram. We have had great success teaching Bayes' Rule
with this method for several years and received positive feedback via class instruction surveys from our students when incorporating this material in conjunction with the standard mathematical presentation of Bayes' Rule as given in Equation \ref{eq1}.

\section{Acknowledgments}

The authors thank and appreciate the constructive comments and suggestions made over the years regarding the visual approach of instructing Bayes' Rule. The students struggling with the material made us realize that another approach needed to be adopted in order to teach this fundamental statistical law. Additionally, the authors thank their colleagues for constructive comments.

\bibliographystyle{dcu}

\end{document}